\documentclass[11pt]{article}

\pdfoutput=1
\usepackage{graphicx, hyperref}
\usepackage{amsmath}
\usepackage{xcolor}
\usepackage{cancel}
\usepackage[normalem]{ulem}
\usepackage{amssymb}
\usepackage{slashed}
\usepackage{stmaryrd}
\usepackage[T1]{fontenc}
\usepackage[utf8]{inputenc}
\usepackage[english]{babel}
\usepackage{amsthm}
\usepackage{tikz-cd}
\usepackage{mathabx}
\usepackage{braket}
\usepackage{physics}
\usepackage{geometry}
\usepackage{cite}
\usepackage{mathtools}

\usepackage{mathrsfs}
\numberwithin{equation}{section}

\newcommand{\beq}{\begin{eqnarray}}
\newcommand{\eeq}{\end{eqnarray}}

\DeclareMathOperator{\extdm}{d}
\newcommand{\extd}{\extdm \!}

\hypersetup{
pdfstartview = {FitH},
}
\hypersetup{
	colorlinks=true,         
	linkcolor=black,          
	citecolor=red,        
	urlcolor=blue            
}

\SetSymbolFont{stmry}{bold}{U}{stmry}{m}{n}

\def\me{\mathfrak e}

\newcommand{\sq}{{\sqrt{q}}}
\newcommand{\reff}{{\text{vac}}}
\newcommand{\vac}{{\text{vac}}}

\newenvironment{align1}
{\begin{equation}\begin{aligned}}
{\end{aligned}\end{equation}}

\hypersetup{
pdfstartview = {FitH},
}
\hypersetup{
	colorlinks=true,         
	linkcolor=blue,          
	citecolor=magenta,        
	urlcolor=magenta            
}

\begin{document}

\title{{\Large{\textbf{\sffamily   Teleparallel gravity at null infinity}}} }
\author{\sffamily Florian Girelli${}^1$, Jianhui Qiu${}^{1,2,3}$, Céline Zwikel${}^4$
\date{\small{\textit{
$^1$Department of Applied Mathematics, University of Waterloo,
200 University Avenue West, Waterloo, Ontario, Canada, N2L 3G1\\
\vspace{0.2cm}
$^2$National Astronomical Observatories, Chinese Academy of Sciences,
Beijing 100101, China\\
\vspace{0.2cm}
$^3$School of Astronomy and Space Sciences, University of Chinese Academy of Sciences,
No. 19A, Yuquan Road, Beijing 100049, China\\
\vspace{0.2cm}
$^4$Perimeter Institute for Theoretical Physics, 31 Caroline Street North, Waterloo, Ontario, Canada N2L 2Y5\\}}}}

\maketitle

\begin{abstract}
Four-dimensional asymptotically flat spacetimes have been central to recent developments in infrared physics. Gravitational waves reaching the asymptotic boundary reveal an infinite-dimensional symmetry group known as the Bondi-Metzner-Sachs (BMS) group. The vacuum structure breaks this symmetry, giving rise to Goldstone modes that play a pivotal role in the analysis of scattering amplitudes. However, these modes must be added to the phase space of the metric formulation.

In this work, we explore an alternative formulation of general relativity, teleparallel gravity, which is dynamically equivalent to the standard metric description in the bulk. This framework relies on a tetrad field to encode the gravitational degrees of freedom and a flat connection to represent inertial effects. Leveraging this decomposition, we propose a novel encoding of the Goldstone modes within the flat connection. We examine the implications of this approach in the covariant phase space framework, focusing on the symplectic potential and its connection to the Wald–Zoupas prescription. 
\end{abstract}

\tableofcontents

\section{Introduction}
Symmetries play a key role in defining observables in physics. For gauge theories such as gravity, the generators of symmetries are defined at the boundary. For instance, the electric charge is measured using the Gauss law. 
In this work, we study four-dimensional gravitational spacetimes and their asymptotic boundaries, with particular emphasis on future null infinity (Scri) in the context of asymptotically flat spacetimes.

In the seminal papers \cite{Bondi:1962px,Sachs:1962zza}, the authors exhibited the symmetry group preserving Scri for a choice of boundary conditions. It is an infinite extension of the Poincaré group by supertranslations. This group is sometimes referred to as the historical BMS, and we will adopt this terminology. Recently it was shown that the action of supertranslations precisely corresponds to the displacement memory effect \cite{Strominger:2014pwa} and to the leading soft graviton theorem \cite{Strominger:2013jfa,He:2014laa}. This establishes a deep connection between low-energy phenomena of gravity resulting a resurgence of interest for infrared physics \cite{Strominger:2017zoo}. 
Boundary conditions were proposed yielding extensions of the historical BMS group, accommodating superrotations for extended BMS \cite{Barnich:2009se,Barnich:2010eb,Barnich:2011mi} and generalized BMS \cite{Campiglia:2014yka,Campiglia:2015yka,Compere:2018ylh,Campiglia:2020qvc}. There is also an option to realize Weyl rescalling \cite{Freidel:2021fxf} and additional symmetries by relaxing the Bondi gauge \cite{Geiller:2022vto,Geiller:2024amx,McNees:2025acf,Campoleoni:2023fug}.

An important consequence of BMS symmetries is the degeneracy of the vacuum state. A BMS transformation does not preserve the Minkowski metric, but it preserves its vanishing energy. In \cite{Compere:2016jwb}, the authors explicitly construct the BMS invariant vacua by acting on Minkowski metric with a finite BMS diffeomorphism. These vacua are labelled by Goldstone modes. These modes do transform under BMS symmetries and have to be included in the phase space. Taking into account the vacuum structure is crucial to define a covariant notion of radiation in asymptotically flat spacetimes \cite{Geroch1977,Compere:2018ylh,Rignon-Bret:2024gcx}. The Goldstone mode can be re-interpreted as edge modes, they appear as contribution to render non-covariant quantities, covariant.

From a purely metric perspective the Goldstone modes are additional fields in the phase space. However there are different formulations of gravity with different fundamental fields defining the theory. 
Our main result is that these Goldstone modes are already present when considering the teleparallel formulation of gravity.
We will show that the Goldstone modes are encoded in the Lagrangian through the flat connection, called the Weitzenböck connection. In a consistent manner with the edge mode interpretation, this connection can be seen as boundary data. 

Gravity, as a tensor theory, can be encoded in many geometrical ways. Different representations provide different insights on the nature of gravity  by highlighting different types of symmetries. In the bulk these different formulations are equivalent, but at the boundaries,  they typically differ. This is most relevant to understand when attempting to quantize gravity. Therefore having a full understanding of the different formulations, in particular their specificity  at the boundary (such as at Scri for example)  is important. We will focus here on the teleparallel formulation which has not been fully explored in terms of the asymptotic symmetries, to the best of our knowledge. 

The most common representation is using the metric tensor and a metric compatible connection, the Levi-Civita connection. The latter has no torsion and the presence of gravity is encoded in a non-zero (Riemannian) curvature tensor. The equation of motion, Einstein's equations, are second order. Another well-known formulation is the Einstein-Cartan formalism, where one uses the frame field and the spin connection. The latter can have both torsion and curvature. In this case the equations of motion, the analogue of Einstein equation and the torsionless condition on the connection, are first order. There is finally a third option, the teleparallel one, where one works with the frame field and the Weitzenböck connection. This latter has zero curvature but the presence of gravity is encoded in the torsion. The equations of motion are second order. In fact, just like General Relativity can be recovered on-shell of some equation of motion, teleparallel gravity can  also be recovered by going on-shell of some equations of motion \cite{Dupuis:2019unm} from the Einstein-Cartan formalism. 

One of the conceptual strengths of the teleparallel formulation is the fact that inertia and gravity are actually disentangled from each other, see for example \cite{Aldrovandi:2013wha}. In particular the Weitzenböck connection only encodes the inertial features, ie the arbitrariness of the choice of coordinates, while gravity degrees of freedom are encoded in the contorsion. Mathematically, this is manifest in the decomposition of the Levi-Civita connection, in terms of the contorsion and the  Weitzenböck connection. One another consequence of this decomposition is that the geodesic equation can take the shape of Newton's equation, with the contorsion playing the role of the "gravitational force". 

The Einstein-Hilbert action and the teleparallel action are equivalent, up to a boundary term. We derive the impact of this boundary term on covariant phase methods, which provide a procedure to determine the generators, i.e. the charges, associated to the asymptotic symmetries \cite{GAWEDZKI1972307, Kijowski:1973gi, Kijowski:1976ze, Crnkovic:1986ex, Ashtekar:1990gc, Lee:1990nz, Wald:1993nt, Iyer:1994ys, Wald:1999wa, Barnich:2003xg}. 
These features also occur for different formulations of general relativity \cite{DePaoli:2018erh,Oliveri:2019gvm,Freidel:2020xyx}. {Covariant phase methods were used for teleparallel gravity to derive thermodynamical quantities of stationary solutions such as Kerr and Schwarzschild black holes \cite{Maluf:2002zc,Maluf:2012na,Hammad:2019oyb}. The novelty of our work in to consider gravitational waves and discuss the split between inertia and propagating degrees of freedom in Bondi gauge.}

Importantly, we want to select a symplectic potential yielding  well-defined charges that are physically sound. This is encoded in the Wald-Zoupas (WZ) prescription \cite{Wald:1999wa, Grant:2021sxk,Odak:2022ndm,Rignon-Bret:2024wlu,Rignon-Bret:2024gcx}.
This prescription is based on stationary and covariance criteria such that when computing the time evolution of charges we precisely obtain the physical flux. This can be make possible by the inherent ambiguities in defining the symplectic potential. 
In certain instances, these terms can be interpreted as coming from boundary Lagrangians \cite{Compere:2008us, Detournay:2014fva, Compere:2020lrt, Freidel:2020xyx, Fiorucci:2020xto, deHaro:2000xn, Chandrasekaran:2021vyu, Bianchi:2001kw, Freidel:2021cjp, Margalef-Bentabol:2020teu, G:2021xvv, Margalef-Bentabol:2022zso,Capone:2023roc} or from other components of the symplectic potential \cite{Compere:2018ylh,McNees:2023tus,Riello:2024uvs,McNees:2024iyu}.

Recently the Wald-Zoupas prescription has been carefully applied for historical and extended BMS \cite{Grant:2021sxk,Odak:2022ndm,Rignon-Bret:2024gcx} by identifying a Wald-Zoupas symplectic potential. This amounts to include the Goldstone modes in the symplectic potential. While the latter are absent in a metric formulation of gravity, we show they are  implemented in teleparallel gravity. 
We then discuss the relation of the teleparallel symplectic potential that carries the Goldstone modes with the Wald-Zoupas symplectic potential.

\subsection*{Plan}
We start by reviewing teleparallel gravity in section \ref{sec:tele}. We recall its relation to the Einstein action and construct its symplectic potential. In section \ref{sec:bondi}, we consider a tetrad compatible with the Bondi gauge and review the asymptotic symmetry algebra. We comment on the gravitational degrees of freedom in this gauge. In section \ref{sec:vacuum}, we review the introduction of Goldstone modes used to describe the degeneracy of the vacuum structure of Minkowski spacetime. We then propose a natural way of encoding these Goldstones in the teleparallel action using the flat connection. Finally in section \ref{sec:symplecticpot} we discuss the Wald-Zoupas procedure for the symplectic potential of teleparallel gravity.

\section{Teleparallel gravity}\label{sec:tele}
In this section, we review the teleparallel gravity action, the equation of motion and its symplectic potential.
We refer to the book \cite{Aldrovandi:2013wha} and the review \cite{Bahamonde:2021gfp} for a detailed review of the theory. 

\subsection{Action of teleparallel gravity} 
We first introduce the action of teleparallel gravity and define the relevant quantities. We then relate it to Einstein-Hilbert action and finally we recall one of the main motivation to define teleparallel by considering the geodesic equation. 

The Lagrangian density of teleparallel gravity functionally depends on tetrad $\boldsymbol{e}${,  also called frame field,} and on the (flat) spin connection $\boldsymbol{\omega}$ 
\begin{equation}\label{eq:lagtele}
\mathcal L\left[\boldsymbol{e},\boldsymbol{\omega} \right]=\frac{1}{32 \pi G}e\, T^\rho{}_{\mu \nu} S_\rho{}^{\mu \nu}d^nx,
\end{equation} 
where $G$ is Newton constant, $n$ the spacetime dimension, $e$ the volume element, $T$ the torsion and $S$ the superpotential defined in terms of the torsion and the contorsion, as we are going to recall. The spin connection  is taken to be purely inertial,  this means that  the curvature associated to $\omega$ vanishes. There exists therefore  a local Lorentz transformation $\Lambda$, such that
\begin{equation}
\omega_{\mu}{}^{a}{}_b=(\Lambda^{-1})^a{}_c\partial_\mu \Lambda^c{}_b.
\end{equation} 
From the tetrad $\boldsymbol{e}$ and the spin connection $\boldsymbol{\omega}$, one  builds the Weitzenböck connection
\begin{equation}
    \Gamma^\mu{}_{\nu\rho}=e^\mu_a\partial_\rho e_\nu^a+e^\mu_a\omega_{\rho}{}^{a}{}_be^a_\nu
\end{equation}
where $e^\mu_a$ is the inverse of $e_\mu^b$ such that $e^\mu_ae_\mu^b=\delta_a^b$. The torsion is defined as
\begin{equation}
T^\rho{}{}_{\mu \nu}=\Gamma^\rho{}_{\mu\nu}-\Gamma^\rho{}_{\nu\mu}=e^\rho_a\left( \partial_\mu e_\nu^a-\partial_\nu e_\mu^a+ \omega_{\mu}{}^{a}{}_be^b_\nu- \omega_{\nu}{}^{a}{}_be^b_\mu\right)
\end{equation}
and the contorsion tensor as
\begin{equation}
K^\mu{}_{\nu\rho}=\frac12\left(T_\nu{}^\mu{}_\rho+T_\rho{}^\mu{}_\nu-T^\mu{}_{\nu\rho}\right)\,.
\end{equation}
Note that this tensor is antisymmetric in its two first indices. The contorsion can be used to relate the spin connection to the Levi-Civita connection $\overset{\circ}\omega$ (we will use $\circ$ to encode quantities defined in terms of the Levi-Civita connection)
\begin{equation}\label{LC-W}
 \omega_\mu{}^{a}{}_{b}=    
 \overset{\circ}\omega_\mu{}^{a}{}_{b}+  K_\mu{}^a{}_{b}  
\end{equation}
Finally, the superpotential is 
\begin{equation}\label{def:superpotential}
S_\rho{}^{\mu \nu}=K^{\mu \nu}{}_\rho-\delta_\rho^\nu T^\mu+\delta_\rho^\mu T^\nu
\end{equation}
where \begin{equation}\label{torsion vector}
    T^\mu=T^{\nu\mu}{}_\nu
\end{equation} is referred to as the \textit{torsion vector}.

\medskip 
It is well-known that the Lagrangian \eqref{eq:lagtele} differs from the  Einstein-Hilbert Lagrangian by a boundary term that depends on the torsion \cite{Aldrovandi:2013wha}.  
\begin{equation}\label{eq:teleparralintermsofEH}
\mathcal L\left[\boldsymbol{e},\boldsymbol{\omega} \right]=
\stackrel{\circ}{\mathcal{L}}_{\text{\tiny EH}}-\partial_\mu\left(\frac{1}{8\pi G} e\,T^\mu\left[\boldsymbol{e},\boldsymbol{\omega} \right]\right)d^nx
\end{equation}
where $\stackrel{\circ}{\mathcal{L}}_{\text{\tiny EH}}=-\frac1{16\pi G} \sqrt{-g}R[\overset\circ \omega]d^nx$.  We note that since the Einstein-Hilbert Lagrangian is invariant under local Lorentz transformations, the equality \eqref{eq:teleparralintermsofEH} is true for any choice of $\boldsymbol{\omega}$, in particular when $\boldsymbol{\omega} =0$. Hence we can infer that \cite{Krssak:2015lba}  
\begin{align}\nonumber
&\mathcal L\left[\boldsymbol{e},\boldsymbol{\omega} \right]+\partial_\mu\left[\frac{1}{8\pi G} e\, T^\mu[\boldsymbol{e},\boldsymbol{\omega}]\right] =\mathcal L\left[\boldsymbol{e},0 \right]+\partial_\mu\left[\frac{1}{8\pi G} e\, T^\mu[\boldsymbol{e},0]\right] \\ 
\label{def:tellLagr}
\Leftrightarrow& \mathcal L\left[\boldsymbol{e},\boldsymbol{\omega} \right]=\mathcal L\left[\boldsymbol{e},0 \right]+\partial_\mu\left[\frac{1}{8\pi G} e\,\omega^{\mu} \right]
\end{align}
where $\omega^\mu:=\omega_\nu{}^{\nu\mu}=\omega_\nu{}^{ ab}e_a^\nu e_b^\mu$  comes from difference of the torsion (vector) contributions, 
\begin{equation}\label{def:spinveccon}
    T^\mu[\boldsymbol{e},0]-T^\mu[\boldsymbol{e},\boldsymbol{\omega}] =\omega^{\mu} .
\end{equation} 
 We will often call $\omega^{\mu}$ the \textit{spin connection vector}.   
The key-point is that, in the teleparallel formalism, the connection information is equivalent to some boundary data. In fact, one can check that the Lagrangian  $\mathcal L\left[\boldsymbol{e},0 \right]$ is  invariant under local Lorentz transformations, \textit{only up to a boundary term} \cite{Krssak:2015lba}. This is similar to the Chern-Simons action in three dimensions. 

\medskip 

 One of the main motivations to introduce the teleparallel formalism is to disentangle gravity degrees of freedom from inertia \cite{Aldrovandi:2013wha}. Indeed, the decomposition \eqref{LC-W} allows to rewrite the Levi-Civita connection into a purely inertial component, the Weitzenböck  connection $\boldsymbol{\omega}$ and a component containing purely the gravitational degrees of freedom, the contorsion $K$. 
\begin{equation}\label{geodeq}
    \frac{\extd^2x^\mu}{\extd s^2} + \overset\circ\omega^\mu{}_{\alpha\beta} \frac{\extd x^\alpha}{\extd s } \frac{\extd x^\beta}{\extd s }=0\Leftrightarrow  \frac{\extd^2x^\mu}{\extd s^2} + \omega^\mu{}_{\alpha\beta} \frac{\extd x^\alpha}{\extd s } \frac{\extd x^\beta}{\extd s }= K^\mu{}_{\alpha\beta} \frac{\extd x^\alpha}{\extd s } \frac{\extd x^\beta}{\extd s }. 
\end{equation}

\subsection{Equation of motions and symplectic potential} 
The variation of the Lagrangian density $\mathcal L\left[\boldsymbol{e},\boldsymbol{\omega} \right]$ defined in eq. \eqref{def:tellLagr} is 
$$\delta \mathcal L =
\frac{\delta \mathcal L}{\delta \phi}\delta\phi+\partial_\mu\Theta^\mu[\phi;\delta \phi], \quad \phi=(\boldsymbol e, \boldsymbol \omega).$$ 
The equations of motion are
\begin{align}
\frac{\delta \mathcal L}{\delta e_\mu^a} &=\frac1{8\pi G} \partial_\nu \left( e S_a{}^{\mu\nu} \right)- e\mathcal J^\mu_a  \,, \qquad \frac{\delta \mathcal L}{\delta \omega_{\mu}{}^{a}{}_b}   = 0
\end{align}
where 
\begin{equation}
    \mathcal J^\mu_a[\boldsymbol{e},0] := -\frac{1}{32 \pi G} \left(e^\mu_a  \, T^\rho{}_{\sigma \nu}[\boldsymbol{e},0] S_\rho{}^{\sigma \nu}[\boldsymbol{e},0] +4e_a^\rho S_b{}^{\mu\nu}[\boldsymbol{e},0]T^b{}_{\nu\rho}[\boldsymbol{e},0] \right)
\end{equation}
and the symplectic potential is
\begin{equation}\label{eq:symplpotteleparallel}
    \Theta_{{\text{\tiny //}}}^\mu[\phi;\delta \phi] = \frac1{8\pi G} \left(e\,S_a{}^{\mu\nu}[\boldsymbol{e},0]\delta e^a_\nu  +  \delta \left( e\,\omega^{\mu}\right)\right) =: \Theta^\mu_S [\boldsymbol{e};\delta e] +   \delta \left( e\,\omega^{\mu}\right)  .  
\end{equation}
We emphasize that  the Lagrangian $ \mathcal L\left[\boldsymbol{e},\boldsymbol{\omega} \right]$ has a dependence on $\boldsymbol{\omega}$ only through a boundary term which explains why the variation is simply zero. See also \cite{Golovnev:2017dox, Hohmann:2021fpr} for a discussion regarding the variations of the connection  degrees of freedom. 
In the following, the boundary contribution from $\omega^{\mu}$ will be very important as it will encode the vacuum structure relevant to have a covariant picture when dealing with the Bondi gauge.

\medskip 

The bulk equations of motion derived from Einstein-Hilbert Lagrangian and teleparallel \eqref{def:tellLagr} are equivalent but the symplectic potential for each theory differ by a total derivative $\extd\vartheta[\boldsymbol{e};\boldsymbol{\omega}]$. The Einstein-Hilbert symplectic potential is 
 \begin{align}\label{symplpotEH}
 \Theta_{{\text{\tiny EH}}}^\mu [\boldsymbol{g};\delta g] &= -\frac1{16\pi G}\sqrt{-g}\left(\nabla_\nu(\delta g)^{\mu\nu} -\nabla^\mu (\delta g)_\nu^\nu \right) =-\frac1{8\pi G}\sqrt{-g} g^{\mu[\rho} g^{\nu] \sigma} \nabla_\nu \delta g_{\rho \sigma}
 \end{align}
Using \eqref{eq:teleparralintermsofEH} and \eqref{def:spinveccon}, we have 
 \begin{align}
 \Theta_{{\text{\tiny EH}}}^\mu [\boldsymbol{g};\delta g] &= \Theta_{{\text{\tiny //}}}^\mu[\phi;\delta \phi]  +\frac{1}{8\pi G} \delta(e\,T^\mu\left[\boldsymbol{e},\boldsymbol{\omega} \right]) +\partial_\nu\vartheta^{\mu\nu}[\boldsymbol{e},\boldsymbol{\omega}]  \nonumber \\
 & =   \frac1{8\pi G} \left(e\,S_a{}^{\mu\nu}[\boldsymbol{e},0]\delta e^a_\nu  +  \delta \left( e\,\omega^{\mu}\right)+\delta(e\,T^\mu\left[\boldsymbol{e},\boldsymbol{\omega} \right])\right)+\partial_\nu\vartheta^{\mu\nu}[\boldsymbol{e},\boldsymbol{\omega}] \nonumber\\
 &=\frac1{8\pi G}\left( e\,S_a{}^{\mu\nu}[\boldsymbol{e},0]\delta e^a_\nu + \delta(e\,T^\mu\left[\boldsymbol{e},0\right]) \right)+\partial_\nu\vartheta^{\mu\nu}[\boldsymbol{e},\boldsymbol{\omega}]
 \label{eq:relationbtwsymplecticpotential}
\end{align}
which allows to find what is $\vartheta[\boldsymbol{e},\boldsymbol{\omega}] $, up to an exact form\cite{DePaoli:2018erh,Oliveri:2019gvm}: 
\begin{equation}\label{cornertermEH//}
    \vartheta^{\mu\nu}[\boldsymbol{e},\boldsymbol{\omega}]=  \frac{1}{16\pi G}\left(e^\nu_{ b}\delta e^{b\mu}- e^\mu _b \delta e^{b\nu} \right)\,.
\end{equation}
Note that this is the same corner term that relates the symplectic potential of the tetrad formulation of general relativity to the Einstein-Hilbert symplectic potential.

\section{Bondi gauge for the frame field in a nutshell }\label{sec:bondi}
From now on, we will restrict our study of teleparallel gravity to asymptotic flat spacetimes in four dimensions. We will focus on the future null boundary. We will use the Bondi gauge that is well suited to describe radiation when approaching the asymptotic boundary. We will review the Bondi gauge and discuss the residual symmetries in this gauge. 

\subsection{Bondi gauge in terms of the frame field}
The original Bondi gauge was set up in the metric formalism \cite{Bondi:1962px,Sachs:1962zza}. It uses the Bondi coordinates $(u,r,x^A)$ where $u$ is the retarded time, $r$ the areal distance and $x^A$ the coordinates on the celestial sphere. 
This gauge is defined by the conditions 
\begin{equation}\label{Bondigaugecondition}
g_{rr}=0=g_{rA},\qquad \partial_r(\det \gamma_{AB}/r^4)=0
\end{equation}
where $\gamma_{AB}$ is the transverse metric. 
The line element in Bondi gauge can be written as  
\begin{equation}\label{metricBondi} 
\extd s^2=g_{\mu\nu} \extd x^\mu \extd x^\nu=-2 e^{2 \beta} \extd u(\extd r+F \extd u)+\gamma_{A B}\left(\extd x^A-U^A \extd u\right)\left(\extd x^B-U^B \extd u\right)\,, 
\end{equation}
where $F, \beta$ and $U^A$ are arbitrary functions of $(u,r,x^A)$. Furthermore we choose the coordinates on the celestial sphere to be the stereographic coordinates $(z,\bar z)$ such that the metric round 2-sphere $q_{AB}$ is given by 
\begin{equation}\label{stereocoord}
q_{AB}=\frac{4}{\left(1+z\bar z\right)^2} \extd z\extd \bar z \,.
\end{equation}

The fundamental field in teleparallel gravity is not the metric but the the frame field $\boldsymbol{e}$. We introduce a frame field compatible with \eqref{metricBondi} via
\begin{equation}
 g_{\mu\nu}=e_\mu^a\eta_{ab}e_\nu^b   
\end{equation}
where we take the Minkowski metric to be 
$\eta=-(\extd x^0)^2+(\extd x^1)^2+2(\extd x^2)(\extd x^3)$. Explicitly the tetrad $e_\mu^a$  and its  inverse $e^\mu_a$ are respectively  \begin{align1} \label{tetradBondi}
e^0& = \left(F+\frac{e^{2 \beta}}{2}\right) \extd u +\extd r  \,,\quad
e^1 =\left(F-\frac{e^{2 \beta}}{2}\right)  \extd u +\extd r 
\,,\quad
e^i=rE_A{}^i\left(\extd x^A-U^A \extd u\right), \\ 
E^2&=
\frac{\gamma_{zz}}{\gamma_{z\bar z}+\sqrt{-\gamma}} \,\extd z+\extd \bar z\,, \qquad
E^3=\frac12\left(\gamma_{z\bar z}+\sqrt{-\gamma}\right) \,\extd z+\frac12 \gamma_{\bar z\bar z}\extd \bar z \\
 e_0&=e^{-2\beta} \left(\partial_u +\left(\frac{e^{2\beta}}{2}-F\right)\partial_r +U^A\partial_A \right) \,, \quad e_1= e^{-2\beta} \left(- \partial_u +\left(F +\frac{e^{2\beta}}{2}\right)\partial_r- U^A\partial_A\right) \,,\\ e_i&=\frac{1}{r}E_i^A\partial_A
\end{align1}
with $i=2,3$. 
Note that the tetrad \eqref{tetradBondi} differs from the choice in \cite{Godazgar:2020kqd,Freidel:2021fxf} as we have a different convention for the Minkowski metric.

Finally we impose the fall-off conditions 
\begin{align1}
\label{BMSBCfall}
   \beta&= \frac1{r^2}\beta_2 + O(r^{-3})\,,\qquad U^A=\frac1{r^2} U^A_2 + O(r^{-3})\,, \\\gamma_{AB}&=r^2q_{A B}+r\,C_{AB}+\frac14q_{AB}C_{MN}C^{MN} +O(r^{-1}) 
\end{align1}
where we take ${q}_{A B}$ to be the round sphere metric in stereographic coordinates \eqref{stereocoord}.  The Bondi gauge condition on the determinant imposes that the shear $C_{AB}$ is trace free. We denote $D_A$ the covariant derivative compatible with $q_{AB}$.
The equations of motion impose
\begin{equation}
\begin{aligned}
\beta_2 & = -\frac{1}{32}C_{AB}C^{AB} \\
U^A&=\frac1{r^2} U^A_2 + \frac1{r^3}U^A_3 +O(r^{-4})\,,\quad U^A_2=-\frac12 D_BC^{AB} \\
F&=\frac14 R[q]-\frac{M}{r}+O\left(r^{-2}\right), 
\label{U2}\end{aligned}
\end{equation}
where $M$ is the Bondi mass aspect, $U_3^A$ the angular momentum aspect and $R[q]$ the Ricci scalar associated to $q$ (which is 2 for the 2-sphere).

An important ingredient in teleparalell gravity is the torsion vector \eqref{torsion vector} whose relevant components in Bondi gauge are 
\begin{align}\label{sympltorsion}
e\,T^r\left[\boldsymbol{e},0\right]& = \sq R[q]\,r  - \sq \left(3M+D_AU^A_2\right)+O\left(r^{-1}\right)\,,\qquad
e\,T^u\left[\boldsymbol{e},0\right] = -2\sq \, r +O\left(r^{-1}\right)\,.
\end{align}

\subsection{Residual symmetries in Bondi gauge}
It is well known that the residual symmetries preserving \eqref{Bondigaugecondition} are spanned by a vector $\xi$ whose components satisfy 
\begin{equation}\label{BMSAKV}
   \partial_r \xi^u=0 \,, \quad  \partial_r \xi^A=e^{2\beta}\gamma^{AB}\partial_B\xi^u\,, \quad \xi^r=-\frac12r(D_A\xi^A- U^B\partial_Bf)\,.
\end{equation}
The fall-off conditions \eqref{BMSBCfall} further imposes 
\begin{equation}\label{AKV}
\xi^u=T(x^A)+\frac u2 D_AY^A :=f(u,x^A)\,,\quad Y^A=Y^A(x^A)\,,   
\end{equation}
where $Y^A$ are conformal Killing vectors of the 2-sphere. In stereographic coordinates, this implies that $Y^A\partial_A=\mathcal Y(z)\partial_z+\bar{\mathcal  Y}(\bar z)\partial_{\bar z}$.  The three functions $T(x^A)$ and $(\mathcal Y(z),\bar {\mathcal Y}(\bar z))$ label the residual symmetries in this gauge and satisfy the so-called BMS algebra. 

To be more precise, when we restrict to the globally well-defined vector $Y^A$ we obtain the historical BMS algebra \cite{Bondi:1962px,Sachs:1962zza}. In particular it implies that $\partial_z^3\mathcal Y(z)=0$ and $\partial_{\bar z}^3\bar{\mathcal  Y}(\bar z)=0$. The historical BMS algebra is a semi-direct algebra of supertranslations generated by $T$ with Lorentz transformations $Y^A$. 
If we allow meromorphic functions of the sphere, the algebra is the extended BMS algebra \cite{Barnich:2009se,Barnich:2010eb,Barnich:2011mi,Barnich:2011ct,Barnich:2013axa} which is a semi-direct sum of supertranslation generated by $T$ and superrotations generated by the  holomorphic and antiholomorphic functions $ \mathcal Y$ and $\bar{\mathcal Y}$. Note that we still have $\delta_\xi q_{AB}=0$.

Using the frame field requires to also consider the internal symmetries
\begin{equation}
\delta_{(\xi, \lambda)} e_\mu^a=\mathcal{L}_{\xi} e_\mu^a-\lambda^a{}_b e_\mu^b,
\end{equation}
where $\lambda$ is a infinitesimal Lorentz transformation. To demand that  $\xi$ defined in \eqref{BMSAKV} preserves the Bondi gauge conditions for the frame field \eqref{tetradBondi} completely fixes the six components of the Lorentz transformation $\lambda$,
 \begin{equation}
\begin{aligned}
(\lambda(\xi))^0{}_1& =\partial_r\xi^r \\
(\lambda(\xi))^{2}{}_3 & = E^2{}_{A}( \partial_{\bar z}\xi^A -U^A\partial_{\bar z}\xi^u)
\\
(\lambda(\xi))^{0}{}_i &= E_i{}^{A}\left(\partial_A \xi^r+\frac12(e^{2\beta}+2F) \partial_A \xi^u\right)\\(\lambda(\xi))^{1}{}_i  & =E_i{}^{A}\left(\partial_A \xi^r+\frac12(-e^{2\beta}+2F) \partial_A \xi^u\right)
\end{aligned}
\end{equation}

An important consideration is the transformation of the shear under residual symmetries. This is important as the time derivative of the shear carries information about the radiation. 
We have
\begin{align1}
    \delta_\xi C_{AB}&=(f\partial_u +\mathcal L_Y-\dot f)C_{AB}-2D_{\langle A}\partial_{B \rangle}f\\ \label{deltaC}
    \delta_\xi \dot C_{AB}&=(f\partial_u +\mathcal L_Y)\dot C_{AB}-2D_{\langle A}\partial_{B\rangle}\dot f
\end{align1}
where $\langle,\rangle$ denotes the symmetric trace free part and $\,\dot{}\,$ the  derivative with respect to the retarded time $u$. A striking consequence of these transformation laws is that $C_{AB}=0$ or $\dot C_{AB}=0$ are not invariant under the BMS algebra due to the inhomogenous terms in their transformation laws\footnote{Note that for historical BMS, $\dot C_{AB}$ is tensor while $C_{AB}$ is not. This can be seen from \[2D_{\langle A}\partial_{B\rangle}\dot f=D_{\langle A}\partial_{B\rangle}D_CY^C= \partial_z^3\mathcal Y \extd z \extd z + \partial_{\bar z}^3\bar{\mathcal  Y} \extd \bar z \extd \bar z\] which is zero for historical BMS.}. They are anomalous fields:  their transformation under a diffeomorphism as phase space variables is    different from the Lie derivative. This difference is the \textit{anomaly} and it is measured by the operator $\Delta_\xi$ \cite{Chandrasekaran:2020wwn, Chandrasekaran:2021vyu, Freidel:2021cjp, Adami:2021sko, Odak:2023pga}. For the  field-space scalars (in Bondi gauge) $\Delta_\xi$ is given by 
\begin{equation}\label{anomalydef}
\Delta_\xi=\delta_\xi - L_\xi     \,, \quad L_\xi:=f\partial_u+\mathcal L_Y
\end{equation}
For instance, $\Delta_\xi C_{AB}=-\dot f C_{AB}-2D_{\langle A}\partial_{B \rangle}f$. 
The inhomogeneity of the transformation laws is rooted in the fact that Minkowski metric is \textit{not} invariant under a BMS transformation, yielding a degenerate vacuum. We will review how this problem was addressed in the metric formalism \cite{Compere:2016jwb,Compere:2018ylh} in section \ref{sec:vacuum} by introducing new fields and by explaining how teleparallel gravity avoids this issue by choosing an appropriate flat spin connection.

\section{Vacuum structure from the Weitzenböck connection}\label{sec:vacuum}

The split between inertia/coordinate arbitrariness and gravity in the teleparallel formulation allows to introduce naturally a "vacuum" structure in the asymptotic analysis. This structure is necessary to have a covariant definition of radiation and vacuum.
In the metric formalism, this was done by the introduction of additional Goldstone bosons to define a BMS invariant vacuum \cite{Compere:2016jwb,Compere:2018ylh}. 
 
For teleparallel gravity, the gravity degrees of freedom, i.e. the information about the curvature, are encoded as expected in $\boldsymbol{e}$. We will use the Weitzenböck connection to encode the information about the vacuum by constructing it as \textit{Minkowski Levi-Civita connection}. Indeed, in the absence of curvature, we can use the the Levi--Civita connection to define the Weitzenböck connection. 

We first review the results of \cite{Compere:2016jwb,Compere:2018ylh}, before using it to construct the appropriate  Weitzenböck  connection.

\subsection{BMS covariant vacuum}
The Minkowski metric is not invariant under BMS symmetries. A BMS transformation does not change the energy of Minkowski spacetime, so it maps the vacuum into a vacuum, but the new solution possesses other non-trivial BMS charges making it physically distinguishable from the original Minkowski. We can label these different vacua by introducing new fields, the Goldstone bosons $C$ and $\Phi$, that transform under BMS. 
The vacua are given by  \cite{Compere:2016jwb,Compere:2018ylh}
\begin{equation}\label{vacmetric}
ds^2=\begin{pmatrix}
-1-\frac{\partial_uU}{\rho}&-\frac{r}{\rho}&\frac12D^BC^{\text{vac}}_{AB}-D_A\rho\\
-\frac{r}{\rho}&0&0\\
\frac12D^BC^{\text{vac}}_{AB}-D_A\rho&0& (r^2+2U_{\text{vac}})q_{AB}+C^\reff_{AB}\,\rho\\
\end{pmatrix}
\end{equation}
where $\rho=\sqrt{r^2+U_\vac}$, $U_{\text{vac}}=\frac18 C^{\text{vac}}_{AB}C_{\text{vac}}^{AB}$
and 
\begin{equation}\label{eq:Cvac}
    C^{\text{vac}}_{AB}=C_{AB}^{(0)}+(u+C)N^{\text{vac}}_{AB}
\end{equation}
with 
\begin{equation}
C_{AB}^{(0)}=-2D_{<A}D_{B>}C(z,\bar z)\,,\quad N^{\text{vac}}_{AB}=\frac12\partial_{<A}\Phi \partial_{B>}\Phi-D_{<A}\partial_{B>}\Phi,
\end{equation}
\begin{equation}
    \Phi=\phi(z)+\bar\phi(\bar z)-\log\left( \frac2{(1+z\bar z)^2} \right).
\end{equation}
The transformation laws of $\Phi$ and $C$ under \eqref{AKV} are
\begin{align}
    \delta_\xi C& = T+Y^A\partial_A C -\frac12 C D_AY^A \,,\quad \delta_\xi\Phi= Y^A\partial_A \Phi+ D_AY^A.
\end{align}
We denote by $\me$ the frame field \eqref{tetradBondi} evaluated on the vacua solution \eqref{vacmetric}.

The Goldstone modes can be used to define BMS covariant radiation. Indeed the $N^{\text{vac}}_{AB}$ has precisely the same property of the tracefree part of the Geroch tensor \cite{Geroch1977}. If we define the news as
\begin{equation}\label{news}
    N_{AB}:=-\dot C_{AB}+ N^{\text{vac}}_{AB}, 
\end{equation}
it transforms homogeneously 
\begin{equation}
    \delta_\xi  N_{AB}=(f\partial_u+\mathcal L_Y) N_{AB}.
    \end{equation}
See for instance section 2 of \cite{Rignon-Bret:2024gcx} for more details on the definition of radiation in asymptotically flat spacetimes.

\subsection{The Weitzenböck connection encodes the vacuum data}
As we recalled earlier, the Weitzenböck connection $\omega$ is flat.  We can then naturally relate it to a "vacuum structure" associated to a reference frame-field $\me$ (containing no curvature), associated with the vacuum metric \eqref{vacmetric}. Indeed we can take the Weitzenböck connection as a flat Levi-Civita connection for $\me$ on the boundary such that  
\begin{equation}\label{vacflatconnection}
   \omega_{\reff}{}_\mu{}_{ab}[\me]:= \overset{\circ}\omega_{\mu ab}[\me]= \me_{[a}^\rho \me_{b]}^\sigma\left(\me_{\sigma c} \partial_\mu \me_\rho^c+\partial_\sigma (\me^a_\rho\me_{a\mu})\right),
\end{equation}
and 
$\omega_{\reff}[\me]$ is flat by construction. 
The vector connection $\omega_{\reff}^\mu:=\omega_{\reff}{}_\nu{}^{\nu\mu}[\boldsymbol{e},\me]= e_a^\nu e_b^\mu\, \omega_{\reff}{}_\nu{}^{ ab}[\me]$ arising as a boundary contribution is built from \textit{both} the vacuum frame field $\me$ \textit{and} the frame field $\boldsymbol{e}$ encoding the  radiation data. 
We have in particular
\begin{align}\label{rspinvector}
e\,\omega^r_{\text{vac}}&=2\sq\,r +\sq \left(-2\,M+\frac{1}2D_AD_BC^{AB}_{\text{vac}}+\frac{1}{4}C_{AB}N^{AB}_{\text{vac}}-\frac{1}{8}C^{AB}_{\text{vac}}N_{AB}^{\text{vac}}\right)+O(r^{-1}),\nonumber \\
e\,\omega^u_{\text{vac}}&=-2\sq\,r +O(r^{-1})\,,\quad e\,\omega^A_{\text{vac}}=\sq\,\partial^A\ln\sq+O(r^{-1}).
\end{align}

\medskip

Finally with this choice of Weitzenböck connection, we can write the Lagrangian density of teleparallel gravity  \eqref{def:tellLagr}
\begin{equation}
\mathcal L\left[\boldsymbol{e},0 \right]+\partial_\mu\left[\frac{1}{8\pi G} e\,\omega_{\reff}^{\mu} \right] =\mathcal L\left[\boldsymbol{e},\boldsymbol{\omega_{\reff}} \right]. \label{bdy-bulk-ref}
\end{equation}
We emphasize that $T^a{}_{\mu\nu}(\boldsymbol{e},\boldsymbol{\omega_\reff})\neq 0$ in general, since $\boldsymbol{e}\neq \me$. 

We note that while the Einstein-Hilbert Lagrangian density has \textit{no} anomaly $\Delta_\xi (\stackrel{\circ}{\mathcal{L}}_{\text{\tiny EH}})=0$, the Lagrangian density of teleparallel gravity has one. Indeed we derive from \eqref{eq:teleparralintermsofEH} that
\begin{equation}\label{anomaly telep}
\Delta_\xi \mathcal L\left[\boldsymbol{e},\boldsymbol{\omega_{\reff}}\right]= \extd a_\xi
\quad \textrm{ with } 
\underset{\leftarrow}{a_\xi}= - \lim_{r\to \infty}
\Delta_\xi \left(\frac{1}{8\pi G} e\,T^r\left[\boldsymbol{e},\boldsymbol{\omega_{\reff}} \right]\right)
\end{equation}
where the symbol $\leftarrow$ under a quantity indicates the pullback of that quantity onto the boundary.

\section{Symplectic potential for teleparallel gravity}\label{sec:symplecticpot}

\subsection{Symplectic potentials and Wald-Zoupas prescription}
In the covariant phase space formalism, a key ingredient for the computation of charges associated to asymptotic symmetries is the symplectic potential. 
Its derivation from an action is not unique when we only impose the bulk equations of motion. We rather obtain an equivalent class of symplectic potentials
\begin{equation}
    \Theta' \sim \Theta+\extd \vartheta +\delta \ell  
\end{equation}
We encountered a first example of this in section \ref{sec:tele} where we related the symplectic potential of teleparallel gravity to the Einstein-Hilbert symplectic potential in \eqref{eq:relationbtwsymplecticpotential}. Importantly, the choice of $\vartheta$ and $\ell$ generically impact the computations of the charges associated to asymptotic symmetries. 

The symplectic potential needs to be selected based on criteria beyond the bulk equations of motion. For instance for closed systems, we require to impose that $\Theta=0$ when imposing boundary equations of motion (Neumann type of boundary conditions) or when imposing chosen boundary conditions (Dirichlet type of boundary conditions). However for leaky boundaries, such as future null infinity, the symplectic potential can not be put to zero without suppressing  radiation. This can be seen from the Ashtekar-Streubel symplectic potential \cite{Ashtekar:1981bq}
\begin{equation}\label{ASsymplpot}
\Theta^{\text{\tiny AS}}=-\frac1{32\pi G}\sq\,\dot C^{AB}\delta C_{AB}\,.
\end{equation}
This motivated Wald and Zoupas\footnote{Note that a potential drawback is that this requirement may be too strong for certain situations. For instance for generalized BMS \cite{Campiglia:2014yka,Campiglia:2015yka,Compere:2018ylh,Campiglia:2020qvc}. However it is achievable for the symmetry algebras we are considering.} to suggest the following prescription \cite{Wald:1999wa} (we have borrowed the recent itemized version of this argument from \cite{Odak:2022ndm}): 
\begin{enumerate}
    \item the symplectic potential  has to be a potential for the pull-back of the symplectic 2-form on the boundary,
    \item it has to be built  from a local and covariant functional of the dynamical fields and background structure,
    \item it has to vanish for conservative boundary conditions, i.e. in absence of radiation. 
\end{enumerate}
To summarize it has to be given by 
\begin{equation}\label{WZsymplpot}
   \Theta^{\text{\tiny WZ}}= \underset{\leftarrow}{\Theta}+\delta b
\end{equation}
where $\Theta^{\text{\tiny WZ}}$ vanishes in absence of radiation and has no anomaly ($\Delta_\xi \Theta^{\text{\tiny WZ}}=0$). Furthermore the authors in \cite{Odak:2022ndm} showed that if we want to define the WZ charges via improved Noether charges, we need to require that $b$ is obtained from a covariant boundary Lagrangian (i.e. there would be no anomaly).

In the rest of the section, we start from Einstein-Hilbert symplectic potential and review the construction of the WZ symplectic potential for Scri where $b$ is anomalous and can be treated as done in \cite{Odak:2022ndm}. We then show how teleparallel theory directly provides a non-anamolous $b$ and therefore yields more directly to the WZ symplectic potential.

\subsection{Einstein-Hilbert symplectic potential}

The pull back of Einstein-Hilbert symplectic potential \eqref{symplpotEH}  on future null infinity in Bondi gauge is 
\begin{align}\label{symplpotEHBondi}
\underset{\leftarrow}{\Theta_{\text{\tiny EH}}}&=-\frac1{32\pi G}\sq\,\dot C^{AB}\delta C_{AB}-\frac{\sq}{16\pi G}\left(2\delta M+ \delta\left({D}_A {U}_2^A +2\partial_u\beta_2\right)\right)
\end{align}
It can be written as \eqref{WZsymplpot} with 
\begin{equation}
\Theta^{\text{\tiny WZ}}=  -\frac1{32\pi G}\sq\,\dot C^{AB}\delta C_{AB} \,,\quad b_{\text{\tiny EH}}= \frac{\sq}{16\pi G}\left(2 M+ {D}_A {U}_2^A  -\frac18 \dot C_{AB}C^{AB}\right)
\end{equation}
so it satisfies criterion 1. For historical BMS, no radiation is defined as $\dot C^{AB}=0$ and $\Delta_\xi( \Theta^{\text{\tiny WZ}})=0$, which makes criteria 2 and 3 to be satisfied. 

Then if we want the WZ symplectic potential to give the improved Noether charges, we need to consider the anomaly of $b$. We write it  for historical and extended BMS and recall that $\xi$ is defined in \eqref{AKV}. For a given vector $V^A$, we have 
\begin{subequations}
\begin{align}
   \Delta_\xi \sq&=-3\dot f \,\sq \\
\Delta_\xi \left({\sq} \dot C_{AB}C^{AB}\right)&=-2\sq\partial_u\left( C^{AB}D_A\partial_Bf \right)\\
   \Delta_\xi\left(\sq\, M\right)&=\sq\left(\frac12D_A\left(N^{AB}\partial_Bf\right)+\frac12C^{AB}D_A\partial_B\dot f -\frac14 \partial_u\left(C^{AB}D_A\partial_Bf\right)\right)\\ \label{Simonetrick}
\Delta_\xi\left(\sq\, D_AV^A\right)&=\sq D_A\left(\left(\Delta_\xi-3\dot f\right) V^A\right)+\sq \partial_u(V^A \partial_Af) 
\end{align}
\end{subequations}
If we take  $V^A=U^A_2$ defined in \eqref{U2}, we have (dropping the total derivative on the sphere) $$\Delta_\xi \left(\sq D_AV^A\right)=\Delta_\xi\left(\sq\, D_AU^A_2\right)=\frac{\sq}2 \partial_u(C^{AB} D_B\partial_Af) $$
For the specific case of historical BMS, the anomaly of $b_{\text{\tiny EH}}$ is \cite{Odak:2022ndm}
\begin{equation}\label{anomalyEH}
\Delta_\xi b_{\text{\tiny EH}}= \frac{\sq}{16\pi G} \frac{1}4\partial_u\left( C^{AB}D_A\partial_Bf \right).
\end{equation}
where we dropped the covariant derivatives on the sphere and used that $C^{AB}D_A\partial_BD_CY^C\, \propto\,  C_{zz} \partial_z^3\mathcal Y(z)+C_{\bar z\bar z} \partial_{\bar z}^3\bar{\mathcal  Y}(\bar z) $ which is zero for historical BMS.
Therefore to be able to reach $\Theta^{\text{\tiny WZ}}$, the authors in \cite{Odak:2022ndm} introduced an anomaly free Lagrangian $\ell^c$ such that $\ell^c=b+dc$. Adding this Lagrangian shifts $\underset{\leftarrow}{\Theta_{\text{\tiny EH}}}\to \underset{\leftarrow}{\Theta_{\text{\tiny EH}}}-\delta \ell^c+d\delta c$ which is precisely $\Theta^{\text{\tiny WZ}}$. 

However adding this boundary Lagrangian has an impact on the improved Noether charges and they check it gives the physically sound  charges. An important aspect of their work is that they work out the boundary Lagrangian \textit{in the Bondi gauge}. They do not provide an expression in terms of the fields of the theory $(\boldsymbol{e},\boldsymbol{\omega} )$. As we shall see teleparallel gravity provides a boundary Lagrangian not tied to a specific gauge, since it is expressed in terms of the fields of the theory (at least for historical BMS).  

\subsection{Teleparallel symplectic potential }
We  evaluate the teleparallel symplectic potential \eqref{eq:symplpotteleparallel} in the Bondi gauge. First we express the superpotential contribution $\Theta^\mu_S[\boldsymbol{e},\delta e]=\frac{1}{8\pi G} e\,S_a{}^{\mu\nu}[\boldsymbol{e},0]\delta e^a_\nu$, 
\begin{align}\label{symplpotsuperpotential}
\Theta^r_S& =\frac{\sq}{16\pi G}\left( 4\delta M -\frac12 \dot C^{AB}\delta C_{AB} \right)+O\left(r^{-1}\right)\,,\qquad
\Theta^u_S= O\left(r^{-1}\right)\,, \qquad \Theta^A_S=O\left(r^{-1}\right)\,.
\end{align}
Interestingly, $\Theta^\mu_S$ has contribution from the contorsion and the torsion \eqref{def:superpotential}. The news terms is purely coming from the contorsion, which encodes the gravitational degrees of freedom in the geodesic equation \eqref{geodeq}. Combining the boundary contribution \eqref{rspinvector} defined in terms of the vacuum field $\me$ and the teleparallel symplectic potential \eqref{symplpotsuperpotential}, the pull back of the symplectic potential of teleparallel gravity \eqref{eq:symplpotteleparallel} in the Bondi gauge  is 
\begin{align}\label{symplpottele}
\underset{\leftarrow}{\Theta_{{\text{\tiny //}}}}&=\frac{\sq}{  16\pi G}\left[ -\frac12 \dot C^{AB}\delta C_{AB}  +\delta \left(D_AD_BC^{AB}_{\text{vac}}+\frac{1}{2}C^{AB}N^{\text{vac}}_{AB}-\frac{1}{4}C^{AB}_{\text{vac}}N^{\text{vac}}_{AB}\right) \right]\,.
\end{align}

As we recalled in section \ref{sec:tele}, the relation between the Einstein-Hilbert action and teleparallel gravity is given by \eqref{eq:relationbtwsymplecticpotential}. Using \eqref{sympltorsion}, \eqref{symplpotsuperpotential} and computing \eqref{cornertermEH//} in Bondi gauge 
 \begin{align}
    16\pi G\, \vartheta^{ur}&=\delta\left(D_AU^A_2-\partial_u\beta_2\right)+O\left(r^{-1}\right).     
\end{align}
we verify that we recover  the Einstein-Hilbert symplectic potential computed in the  Bondi gauge \eqref{symplpotEHBondi}.

We now discuss the teleparallel symplectic potential \eqref{symplpottele} and the WZ criteria. 
For historical BMS, \eqref{symplpottele} can be written as
\begin{equation}
 \Theta^{{\text{\tiny WZ}}}=  -\frac1{32\pi G}\sq\,\dot C^{AB}\delta C_{AB} =\Theta^{{\text{\tiny AS}}}\,,\quad b_{{\text{\tiny //}}}= -\frac{\sq}{16\pi G}\left(D_AD_BC^{AB}_{\text{vac}} \right)
\end{equation}
By using \eqref{Simonetrick}, we have for historical BMS  (dropping the total derivative on the sphere)
\begin{equation}
    \Delta_\xi b_{{\text{\tiny //}}} = 0 \,.
\end{equation}
This means that teleparallel gravity gives for free a non-anomalous symplectic potential. Therefore there is no need to add by hand a boundary Lagrangian to cancel the anomaly! 

When considering the improved Noether charges, although the symplectic potential does not exhibit an anomaly, the teleparallel gravity action is anomalous (in contrast to the covariant Einstein-Hilbert action) and will impact the charges computation. In a sense, we have shifted the anomaly from the symplectic potential to the Lagrangian itself. Indeed, we can explicitly evaluate the Lagrangian anomaly of teleparallel gravity in \eqref{anomaly telep}. We have \begin{equation}
\lim_{r\to \infty}(eT^r\left[\boldsymbol{e},\boldsymbol{\omega_{\reff}} \right])=\sq\left( -M+\frac12D_AD_BC^{AB}-\frac12 D_AD_BC^{AB}_{\text{vac}}+\frac18N^{AB}_{\text{vac}}C_{AB}^{\text{vac}}-\frac14N^{AB}_{\text{vac}}C_{AB}\right)
\end{equation}
and therefore for historical BMS 
$    \Delta_\xi(\lim_{r\to \infty}e T^r)=  -\frac14 \partial_u(C^{AB}D_B\partial_Af)
$
yielding
\begin{equation}
a_\xi=\frac{\sq}{16\pi G} \frac12 \partial_u(C^{AB}D_B\partial_Af)    
\end{equation}
Consistently the improved Noether charges will be impacted by this Lagrangian anomaly, see the precise way in \cite{Odak:2022ndm} in order to recover the same physically sound charges.

To conclude our comparison of symplectic potentials between Einstein-Hilbert gravity and teleparallel gravity, we have demonstrated that, at the level of the action, Einstein-Hilbert gravity is covariant, whereas teleparallel gravity is not. Conversely, at the level of the symplectic potential, teleparallel gravity with the flat connection chosen as in \eqref{vacflatconnection} is covariant for historical BMS, while Einstein-Hilbert gravity is not.  

\section{Conclusion}
Gravity can be characterized in many different but equivalent ways in the bulk: standard metric general relativity,   the teleparallel formulation, Einstein-Cartan formulation,... Some formulation might be more adequate than other to describe the physical situation at hand. The teleparallel formulation uses the frame field (tetrad) to encode the gravitational degrees of freedom and a \textit{flat} connection is used to encode the frame/coordinate system arbitrariness. 

\smallskip

In asymptotically flat spacetimes in Bondi gauge, defining covariant quantities requires to keep track of the vacuum structure, which can be parametrized by Goldstone modes. In the metric formulation, this vacuum structure is added by hand. 
We have shown that the teleparallel formulation encodes these Goldstone modes directly in the Lagrangian, through the flat connection. This suggests that teleparallel gravity could offer a promising framework for encoding a degenerate vacuum beyond the case of null infinity studied here.

\smallskip

Another advantage of teleparallel gravity is that, for the historical BMS group, it yields a Wald--Zoupas--compatible symplectic potential that, unlike the Einstein-Hilbert action, is free from anomalies. It would be interesting to repeat the analysis for enhanced BMS groups, such as the extended BMS group, where a WZ-compatible symplectic potential has been proposed \cite{Rignon-Bret:2024gcx}. We leave this for further investigations.

\section*{Acknowledgments}
We would like to thank Simone Speziale for interesting discussions and comments on the draft. Jianhui Qiu   was supported by the China Scholarship  Project No. 202204910348. Additionally,
research at Perimeter Institute is supported in part by the Government of Canada through the Department of Innovation, Science and Economic Development Canada and by the Province of Ontario through the Ministry of Colleges and Universities.

\providecommand{\href}[2]{#2}\begingroup\raggedright\endgroup


\begin{thebibliography}{10}

\bibitem{Bondi:1962px}
H.~Bondi, M.~G.~J. van~der Burg and A.~W.~K. Metzner, \emph{{Gravitational
  waves in general relativity. 7. Waves from axisymmetric isolated systems}},
  \href{http://dx.doi.org/10.1098/rspa.1962.0161}{\emph{Proc. Roy. Soc. Lond.}
  {\bfseries A269} (1962) 21--52}.

\bibitem{Sachs:1962zza}
R.~Sachs, \emph{{Asymptotic symmetries in gravitational theory}},
  \href{http://dx.doi.org/10.1103/PhysRev.128.2851}{\emph{Phys. Rev.}
  {\bfseries 128} (1962) 2851--2864}.

\bibitem{Strominger:2014pwa}
A.~Strominger and A.~Zhiboedov, \emph{{Gravitational Memory, BMS
  Supertranslations and Soft Theorems}},
  \href{http://dx.doi.org/10.1007/JHEP01(2016)086}{\emph{JHEP} {\bfseries 01}
  (2016) 086}, [\href{https://arxiv.org/abs/1411.5745}{{\ttfamily 1411.5745}}].

\bibitem{Strominger:2013jfa}
A.~Strominger, \emph{{On BMS Invariance of Gravitational Scattering}},
  \href{http://dx.doi.org/10.1007/JHEP07(2014)152}{\emph{JHEP} {\bfseries 07}
  (2014) 152}, [\href{https://arxiv.org/abs/1312.2229}{{\ttfamily 1312.2229}}].

\bibitem{He:2014laa}
T.~He, V.~Lysov, P.~Mitra and A.~Strominger, \emph{{BMS supertranslations and
  Weinberg's soft graviton theorem}},
  \href{http://dx.doi.org/10.1007/JHEP05(2015)151}{\emph{JHEP} {\bfseries 05}
  (2015) 151}, [\href{https://arxiv.org/abs/1401.7026}{{\ttfamily 1401.7026}}].

\bibitem{Strominger:2017zoo}
A.~Strominger, \emph{{Lectures on the Infrared Structure of Gravity and Gauge
  Theory}},  [\href{https://arxiv.org/abs/1703.05448}{{\ttfamily 1703.05448}}].

\bibitem{Barnich:2009se}
G.~Barnich and C.~Troessaert, \emph{{Symmetries of asymptotically flat 4
  dimensional spacetimes at null infinity revisited}},
  \href{http://dx.doi.org/10.1103/PhysRevLett.105.111103}{\emph{Phys. Rev.
  Lett.} {\bfseries 105} (2010) 111103},
  [\href{https://arxiv.org/abs/0909.2617}{{\ttfamily 0909.2617}}].

\bibitem{Barnich:2010eb}
G.~Barnich and C.~Troessaert, \emph{{Aspects of the BMS/CFT correspondence}},
  \href{http://dx.doi.org/10.1007/JHEP05(2010)062}{\emph{JHEP} {\bfseries 05}
  (2010) 062}, [\href{https://arxiv.org/abs/1001.1541}{{\ttfamily 1001.1541}}].

\bibitem{Barnich:2011mi}
G.~Barnich and C.~Troessaert, \emph{{BMS charge algebra}},
  \href{http://dx.doi.org/10.1007/JHEP12(2011)105}{\emph{JHEP} {\bfseries 12}
  (2011) 105}, [\href{https://arxiv.org/abs/1106.0213}{{\ttfamily 1106.0213}}].

\bibitem{Campiglia:2014yka}
M.~Campiglia and A.~Laddha, \emph{{Asymptotic symmetries and subleading soft
  graviton theorem}},
  \href{http://dx.doi.org/10.1103/PhysRevD.90.124028}{\emph{Phys. Rev. D}
  {\bfseries 90} (2014) 124028},
  [\href{https://arxiv.org/abs/1408.2228}{{\ttfamily 1408.2228}}].

\bibitem{Campiglia:2015yka}
M.~Campiglia and A.~Laddha, \emph{{New symmetries for the Gravitational
  S-matrix}}, \href{http://dx.doi.org/10.1007/JHEP04(2015)076}{\emph{JHEP}
  {\bfseries 04} (2015) 076},
  [\href{https://arxiv.org/abs/1502.02318}{{\ttfamily 1502.02318}}].

\bibitem{Compere:2018ylh}
G.~Comp\`ere, A.~Fiorucci and R.~Ruzziconi, \emph{{Superboost transitions,
  refraction memory and super-Lorentz charge algebra}},
  \href{http://dx.doi.org/10.1007/JHEP11(2018)200}{\emph{JHEP} {\bfseries 11}
  (2018) 200}, [\href{https://arxiv.org/abs/1810.00377}{{\ttfamily
  1810.00377}}].

\bibitem{Campiglia:2020qvc}
M.~Campiglia and J.~Peraza, \emph{{Generalized BMS charge algebra}},
  \href{http://dx.doi.org/10.1103/PhysRevD.101.104039}{\emph{Phys. Rev. D}
  {\bfseries 101} (2020) 104039},
  [\href{https://arxiv.org/abs/2002.06691}{{\ttfamily 2002.06691}}].

\bibitem{Freidel:2021fxf}
L.~Freidel, R.~Oliveri, D.~Pranzetti and S.~Speziale, \emph{{The Weyl BMS group
  and Einstein\textquoteright{}s equations}},
  \href{http://dx.doi.org/10.1007/JHEP07(2021)170}{\emph{JHEP} {\bfseries 07}
  (2021) 170}, [\href{https://arxiv.org/abs/2104.05793}{{\ttfamily
  2104.05793}}].

\bibitem{Geiller:2022vto}
M.~Geiller and C.~Zwikel, \emph{{The partial Bondi gauge: Further enlarging the
  asymptotic structure of gravity}},
  \href{http://dx.doi.org/10.21468/SciPostPhys.13.5.108}{\emph{SciPost Phys.}
  {\bfseries 13} (2022) 108},
  [\href{https://arxiv.org/abs/2205.11401}{{\ttfamily 2205.11401}}].

\bibitem{Geiller:2024amx}
M.~Geiller and C.~Zwikel, \emph{{The partial Bondi gauge: Gauge fixings and
  asymptotic charges}},
  \href{http://dx.doi.org/10.21468/SciPostPhys.16.3.076}{\emph{SciPost Phys.}
  {\bfseries 16} (2024) 076},
  [\href{https://arxiv.org/abs/2401.09540}{{\ttfamily 2401.09540}}].


\bibitem{McNees:2025acf}
R.~McNees and C.~Zwikel,
``(Anti)-de Sitter with leaky boundaries and corners,'' [\href{https://arxiv.org/abs/2512.03170}{{\ttfamily 2512.03170}}].

\bibitem{Campoleoni:2023fug}
A.~Campoleoni, A.~Delfante, S.~Pekar, P.~M. Petropoulos, D.~Rivera-Betancour
  and M.~Vilatte, \emph{{Flat from anti de Sitter}},
  \href{http://dx.doi.org/10.1007/JHEP12(2023)078}{\emph{JHEP} {\bfseries 12}
  (2023) 078}, [\href{https://arxiv.org/abs/2309.15182}{{\ttfamily
  2309.15182}}].

\bibitem{Compere:2016jwb}
G.~Comp\`ere and J.~Long, \emph{{Vacua of the gravitational field}},
  \href{http://dx.doi.org/10.1007/JHEP07(2016)137}{\emph{JHEP} {\bfseries 07}
  (2016) 137}, [\href{https://arxiv.org/abs/1601.04958}{{\ttfamily
  1601.04958}}].

\bibitem{Geroch1977}
R.~Geroch, \emph{Asymptotic Structure of Space-Time}, pp.~1--105.
\newblock Springer US, Boston, MA, 1977.

\bibitem{Rignon-Bret:2024gcx}
A.~Rignon-Bret and S.~Speziale, \emph{{Centerless BMS charge algebra}},
  \href{http://dx.doi.org/10.1103/PhysRevD.110.044050}{\emph{Phys. Rev. D}
  {\bfseries 110} (2024) 044050},
  [\href{https://arxiv.org/abs/2405.01526}{{\ttfamily 2405.01526}}].

\bibitem{Dupuis:2019unm}
M.~Dupuis, F.~Girelli, A.~Osumanu and W.~Wieland, \emph{{First-order
  formulation of teleparallel gravity and dual loop gravity}},
  \href{http://dx.doi.org/10.1088/1361-6382/ab7bb7}{\emph{Class. Quant. Grav.}
  {\bfseries 37} (2020) 085023},
  [\href{https://arxiv.org/abs/1906.02801}{{\ttfamily 1906.02801}}].

\bibitem{Aldrovandi:2013wha}
R.~Aldrovandi and J.~G. Pereira, \emph{{Teleparallel Gravity}: {An
  Introduction}}, vol.~173.
\newblock Springer, 2013,
  \href{http://dx.doi.org/10.1007/978-94-007-5143-9}{10.1007/978-94-007-5143-9}.

\bibitem{GAWEDZKI1972307}
K.~GawĘdzki, \emph{On the geometrization of the canonical formalism in the
  classical field theory},
  \href{http://dx.doi.org/https://doi.org/10.1016/0034-4877(72)90014-6}{\emph{Reports
  on Mathematical Physics} {\bfseries 3} (1972) 307--326}.

\bibitem{Kijowski:1973gi}
J.~Kijowski, \emph{{A finite-dimensional canonical formalism in the classical
  field theory}}, \href{http://dx.doi.org/10.1007/BF01645975}{\emph{Commun.
  Math. Phys.} {\bfseries 30} (1973) 99--128}.

\bibitem{Kijowski:1976ze}
J.~Kijowski and W.~Szczyrba, \emph{{A Canonical Structure for Classical Field
  Theories}}, \href{http://dx.doi.org/10.1007/BF01608496}{\emph{Commun. Math.
  Phys.} {\bfseries 46} (1976) 183--206}.

\bibitem{Crnkovic:1986ex}
C.~Crnkovic and E.~Witten, \emph{Covariant description of canonical formalism
  in geometrical theories},  in \emph{Three hundred years of gravitation}
  (S.~Hawking and W.~Israel, eds.), pp.~676--684.
\newblock Cambridge University Press, Cambridge, 1987.

\bibitem{Ashtekar:1990gc}
A.~Ashtekar, L.~Bombelli and O.~Reula, \emph{The covariant phase space of
  asymptotically flat gravitational fields},  in \emph{Mechanics, Analysis and
  Geometry: 200 Years After Lagrange} (M.~Francaviglia, ed.), North-Holland
  Delta Series, pp.~417--450.
\newblock Elsevier, Amsterdam, 1991.
\newblock
  \href{http://dx.doi.org/https://doi.org/10.1016/B978-0-444-88958-4.50021-5}{DOI}.

\bibitem{Lee:1990nz}
J.~Lee and R.~M. Wald, \emph{{Local symmetries and constraints}},
  \href{http://dx.doi.org/10.1063/1.528801}{\emph{J. Math. Phys.} {\bfseries
  31} (1990) 725--743}.

\bibitem{Wald:1993nt}
R.~M. Wald, \emph{{Black hole entropy is the Noether charge}},
  \href{http://dx.doi.org/10.1103/PhysRevD.48.R3427}{\emph{Phys. Rev. D}
  {\bfseries 48} (1993) R3427--R3431},
  [\href{https://arxiv.org/abs/gr-qc/9307038}{{\ttfamily gr-qc/9307038}}].

\bibitem{Iyer:1994ys}
V.~Iyer and R.~M. Wald, \emph{{Some properties of Noether charge and a proposal
  for dynamical black hole entropy}},
  \href{http://dx.doi.org/10.1103/PhysRevD.50.846}{\emph{Phys. Rev.} {\bfseries
  D50} (1994) 846--864}, [\href{https://arxiv.org/abs/gr-qc/9403028}{{\ttfamily
  gr-qc/9403028}}].

\bibitem{Wald:1999wa}
R.~M. Wald and A.~Zoupas, \emph{{A General definition of 'conserved quantities'
  in general relativity and other theories of gravity}},
  \href{http://dx.doi.org/10.1103/PhysRevD.61.084027}{\emph{Phys. Rev. D}
  {\bfseries 61} (2000) 084027},
  [\href{https://arxiv.org/abs/gr-qc/9911095}{{\ttfamily gr-qc/9911095}}].

\bibitem{Barnich:2003xg}
G.~Barnich, \emph{{Boundary charges in gauge theories: Using Stokes theorem in
  the bulk}}, \href{http://dx.doi.org/10.1088/0264-9381/20/16/310}{\emph{Class.
  Quant. Grav.} {\bfseries 20} (2003) 3685--3698},
  [\href{https://arxiv.org/abs/hep-th/0301039}{{\ttfamily hep-th/0301039}}].

\bibitem{DePaoli:2018erh}
E.~De~Paoli and S.~Speziale, \emph{{A gauge-invariant symplectic potential for
  tetrad general relativity}},
  \href{http://dx.doi.org/10.1007/JHEP07(2018)040}{\emph{JHEP} {\bfseries 07}
  (2018) 040}, [\href{https://arxiv.org/abs/1804.09685}{{\ttfamily
  1804.09685}}].

\bibitem{Oliveri:2019gvm}
R.~Oliveri and S.~Speziale, \emph{{Boundary effects in General Relativity with
  tetrad variables}},
  \href{http://dx.doi.org/10.1007/s10714-020-02733-8}{\emph{Gen. Rel. Grav.}
  {\bfseries 52} (2020) 83},
  [\href{https://arxiv.org/abs/1912.01016}{{\ttfamily 1912.01016}}].

\bibitem{Freidel:2020xyx}
L.~Freidel, M.~Geiller and D.~Pranzetti, \emph{{Edge modes of gravity. Part I.
  Corner potentials and charges}},
  \href{http://dx.doi.org/10.1007/JHEP11(2020)026}{\emph{JHEP} {\bfseries 11}
  (2020) 026}, [\href{https://arxiv.org/abs/2006.12527}{{\ttfamily
  2006.12527}}].

\bibitem{Maluf:2002zc}
J.~W. Maluf, J.~F. da~Rocha-Neto, T.~M.~L. Toribio and K.~H. Castello-Branco,
  \emph{{Energy and angular momentum of the gravitational field in the
  teleparallel geometry}},
  \href{http://dx.doi.org/10.1103/PhysRevD.65.124001}{\emph{Phys. Rev. D}
  {\bfseries 65} (2002) 124001},
  [\href{https://arxiv.org/abs/gr-qc/0204035}{{\ttfamily gr-qc/0204035}}].

\bibitem{Maluf:2012na}
J.~W. Maluf, S.~C. Ulhoa and J.~F. da~Rocha-Neto, \emph{{Gravitational pressure
  on event horizons and thermodynamics in the teleparallel framework}},
  \href{http://dx.doi.org/10.1103/PhysRevD.85.044050}{\emph{Phys. Rev. D}
  {\bfseries 85} (2012) 044050},
  [\href{https://arxiv.org/abs/1202.4995}{{\ttfamily 1202.4995}}].

\bibitem{Hammad:2019oyb}
F.~Hammad, D.~Dijamco, A.~Torres-Rivas and D.~B\'erub\'e, \emph{{Noether charge
  and black hole entropy in teleparallel gravity}},
  \href{http://dx.doi.org/10.1103/PhysRevD.100.124040}{\emph{Phys. Rev. D}
  {\bfseries 100} (2019) 124040},
  [\href{https://arxiv.org/abs/1912.08811}{{\ttfamily 1912.08811}}].

\bibitem{Grant:2021sxk}
A.~M. Grant, K.~Prabhu and I.~Shehzad, \emph{{The Wald\textendash{}Zoupas
  prescription for asymptotic charges at null infinity in general relativity}},
  \href{http://dx.doi.org/10.1088/1361-6382/ac571a}{\emph{Class. Quant. Grav.}
  {\bfseries 39} (2022) 085002},
  [\href{https://arxiv.org/abs/2105.05919}{{\ttfamily 2105.05919}}].

\bibitem{Odak:2022ndm}
G.~Odak, A.~Rignon-Bret and S.~Speziale, \emph{{Wald-Zoupas prescription with
  soft anomalies}},
  \href{http://dx.doi.org/10.1103/PhysRevD.107.084028}{\emph{Phys. Rev. D}
  {\bfseries 107} (2023) 084028},
  [\href{https://arxiv.org/abs/2212.07947}{{\ttfamily 2212.07947}}].

\bibitem{Rignon-Bret:2024wlu}
A.~Rignon-Bret and S.~Speziale, \emph{{Covariance and symmetry algebras}},
  [\href{https://arxiv.org/abs/2403.00730}{{\ttfamily 2403.00730}}].

\bibitem{Compere:2008us}
G.~Compere and D.~Marolf, \emph{{Setting the boundary free in AdS/CFT}},
  \href{http://dx.doi.org/10.1088/0264-9381/25/19/195014}{\emph{Class. Quant.
  Grav.} {\bfseries 25} (2008) 195014},
  [\href{https://arxiv.org/abs/0805.1902}{{\ttfamily 0805.1902}}].

\bibitem{Detournay:2014fva}
S.~Detournay, D.~Grumiller, F.~Sch{\"o}ller and J.~Simon, \emph{{Variational
  principle and 1-point functions in 3-dimensional flat space Einstein
  gravity}},
  \href{http://dx.doi.org/10.1103/PhysRevD.89.084061}{\emph{Phys.Rev.}
  {\bfseries D89} (2014) 084061},
  [\href{https://arxiv.org/abs/1402.3687}{{\ttfamily 1402.3687}}].

\bibitem{Compere:2020lrt}
G.~Comp\`ere, A.~Fiorucci and R.~Ruzziconi, \emph{{The $\Lambda$-BMS$_4$ charge
  algebra}}, \href{http://dx.doi.org/10.1007/JHEP10(2020)205}{\emph{JHEP}
  {\bfseries 10} (2020) 205},
  [\href{https://arxiv.org/abs/2004.10769}{{\ttfamily 2004.10769}}].

\bibitem{Fiorucci:2020xto}
A.~Fiorucci and R.~Ruzziconi, \emph{{Charge algebra in Al(A)dS$_{n}$
  spacetimes}}, \href{http://dx.doi.org/10.1007/JHEP05(2021)210}{\emph{JHEP}
  {\bfseries 05} (2021) 210},
  [\href{https://arxiv.org/abs/2011.02002}{{\ttfamily 2011.02002}}].

\bibitem{deHaro:2000xn}
S.~de~Haro, S.~N. Solodukhin and K.~Skenderis, \emph{Holographic reconstruction
  of spacetime and renormalization in the {AdS/CFT} correspondence},
  {\emph{Commun. Math. Phys.} {\bfseries 217} (2001) 595--622},
  [\href{https://arxiv.org/abs/hep-th/0002230}{{\ttfamily hep-th/0002230}}].

\bibitem{Chandrasekaran:2021vyu}
V.~Chandrasekaran, E.~E. Flanagan, I.~Shehzad and A.~J. Speranza, \emph{{A
  general framework for gravitational charges and holographic
  renormalization}},
  \href{http://dx.doi.org/10.1142/S0217751X22501056}{\emph{Int. J. Mod. Phys.
  A} {\bfseries 37} (2022) 2250105},
  [\href{https://arxiv.org/abs/2111.11974}{{\ttfamily 2111.11974}}].

\bibitem{Bianchi:2001kw}
M.~Bianchi, D.~Z. Freedman and K.~Skenderis, \emph{{Holographic
  Renormalization}}, {\emph{Nucl. Phys.} {\bfseries B631} (2002) 159--194},
  [\href{https://arxiv.org/abs/hep-th/0112119}{{\ttfamily hep-th/0112119}}].

\bibitem{Freidel:2021cjp}
L.~Freidel, R.~Oliveri, D.~Pranzetti and S.~Speziale, \emph{{Extended corner
  symmetry, charge bracket and Einstein\textquoteright{}s equations}},
  \href{http://dx.doi.org/10.1007/JHEP09(2021)083}{\emph{JHEP} {\bfseries 09}
  (2021) 083}, [\href{https://arxiv.org/abs/2104.12881}{{\ttfamily
  2104.12881}}].

\bibitem{Margalef-Bentabol:2020teu}
J.~Margalef-Bentabol and E.~J.~S. Villase\~nor, \emph{{Geometric formulation of
  the Covariant Phase Space methods with boundaries}},
  \href{http://dx.doi.org/10.1103/PhysRevD.103.025011}{\emph{Phys. Rev. D}
  {\bfseries 103} (2021) 025011},
  [\href{https://arxiv.org/abs/2008.01842}{{\ttfamily 2008.01842}}].

\bibitem{G:2021xvv}
J.~F.~B. G., J.~Margalef-Bentabol, V.~Varo and E.~J.~S. Villase\~nor,
  \emph{{Covariant phase space for gravity with boundaries: Metric versus
  tetrad formulations}},
  \href{http://dx.doi.org/10.1103/PhysRevD.104.044048}{\emph{Phys. Rev. D}
  {\bfseries 104} (2021) 044048},
  [\href{https://arxiv.org/abs/2103.06362}{{\ttfamily 2103.06362}}].

\bibitem{Margalef-Bentabol:2022zso}
J.~Margalef-Bentabol and E.~J.~S. Villase\~nor, \emph{{Proof of the equivalence
  of the symplectic forms derived from the canonical and the covariant phase
  space formalisms}},
  \href{http://dx.doi.org/10.1103/PhysRevD.105.L101701}{\emph{Phys. Rev. D}
  {\bfseries 105} (2022) L101701},
  [\href{https://arxiv.org/abs/2204.06383}{{\ttfamily 2204.06383}}].

\bibitem{Capone:2023roc}
F.~Capone, P.~Mitra, A.~Poole and B.~Tomova, \emph{{Phase space renormalization
  and finite BMS charges in six dimensions}},
  \href{http://dx.doi.org/10.1007/JHEP11(2023)034}{\emph{JHEP} {\bfseries 11}
  (2023) 034}, [\href{https://arxiv.org/abs/2304.09330}{{\ttfamily
  2304.09330}}].


\bibitem{Riello:2024uvs}
A.~Riello and L.~Freidel, \emph{{Renormalization of conformal infinity as a
  stretched horizon}},
  \href{http://dx.doi.org/10.1088/1361-6382/ad5cbb}{\emph{Class. Quant. Grav.}
  {\bfseries 41} (2024) 175013},
  [\href{https://arxiv.org/abs/2402.03097}{{\ttfamily 2402.03097}}].

\bibitem{McNees:2023tus}
R.~McNees and C.~Zwikel, \emph{{Finite charges from the bulk action}},
  \href{http://dx.doi.org/10.1007/JHEP08(2023)154}{\emph{JHEP} {\bfseries 08}
  (2023) 154}, [\href{https://arxiv.org/abs/2306.16451}{{\ttfamily
  2306.16451}}].


\bibitem{McNees:2024iyu}
R.~McNees and C.~Zwikel,
\emph{{The symplectic potential for leaky boundaries}},
  \href{http://dx.doi.org/10.1007/JHEP01(2025)049}{\emph{JHEP} {\bfseries 01}
  (2025) 049}, [\href{https://arxiv.org/abs/2408.13203}{{\ttfamily
  2408.13203}}].


\bibitem{Bahamonde:2021gfp}
S.~Bahamonde, K.~F. Dialektopoulos, C.~Escamilla-Rivera, G.~Farrugia, V.~Gakis,
  M.~Hendry et~al., \emph{{Teleparallel gravity: from theory to cosmology}},
  \href{http://dx.doi.org/10.1088/1361-6633/ac9cef}{\emph{Rept. Prog. Phys.}
  {\bfseries 86} (2023) 026901},
  [\href{https://arxiv.org/abs/2106.13793}{{\ttfamily 2106.13793}}].

\bibitem{Krssak:2015lba}
M.~Kr\v{s}\v{s}\'ak, \emph{{Holographic Renormalization in Teleparallel
  Gravity}}, \href{http://dx.doi.org/10.1140/epjc/s10052-017-4621-3}{\emph{Eur.
  Phys. J. C} {\bfseries 77} (2017) 44},
  [\href{https://arxiv.org/abs/1510.06676}{{\ttfamily 1510.06676}}].

\bibitem{Golovnev:2017dox}
A.~Golovnev, T.~Koivisto and M.~Sandstad, \emph{{On the covariance of
  teleparallel gravity theories}},
  \href{http://dx.doi.org/10.1088/1361-6382/aa7830}{\emph{Class. Quant. Grav.}
  {\bfseries 34} (2017) 145013},
  [\href{https://arxiv.org/abs/1701.06271}{{\ttfamily 1701.06271}}].

\bibitem{Hohmann:2021fpr}
M.~Hohmann, \emph{{Variational Principles in Teleparallel Gravity Theories}},
  \href{http://dx.doi.org/10.3390/universe7050114}{\emph{Universe} {\bfseries
  7} (2021) 114}, [\href{https://arxiv.org/abs/2104.00536}{{\ttfamily
  2104.00536}}].

\bibitem{Godazgar:2020kqd}
H.~Godazgar, M.~Godazgar and M.~J. Perry, \emph{{Hamiltonian derivation of dual
  gravitational charges}},
  \href{http://dx.doi.org/10.1007/JHEP09(2020)084}{\emph{JHEP} {\bfseries 20}
  (2020) 084}, [\href{https://arxiv.org/abs/2007.07144}{{\ttfamily
  2007.07144}}].

\bibitem{Barnich:2011ct}
G.~Barnich and C.~Troessaert, \emph{{Supertranslations call for
  superrotations}}, \href{http://dx.doi.org/10.22323/1.127.0010}{\emph{PoS}
  {\bfseries CNCFG2010} (2010) 010},
  [\href{https://arxiv.org/abs/1102.4632}{{\ttfamily 1102.4632}}].

\bibitem{Barnich:2013axa}
G.~Barnich and C.~Troessaert, \emph{{Comments on holographic current algebras
  and asymptotically flat four dimensional spacetimes at null infinity}},
  \href{http://dx.doi.org/10.1007/JHEP11(2013)003}{\emph{JHEP} {\bfseries 11}
  (2013) 003}, [\href{https://arxiv.org/abs/1309.0794}{{\ttfamily 1309.0794}}].

\bibitem{Chandrasekaran:2020wwn}
V.~Chandrasekaran and A.~J. Speranza, \emph{{Anomalies in gravitational charge
  algebras of null boundaries and black hole entropy}},
  \href{http://dx.doi.org/10.1007/JHEP01(2021)137}{\emph{JHEP} {\bfseries 01}
  (2021) 137}, [\href{https://arxiv.org/abs/2009.10739}{{\ttfamily
  2009.10739}}].

\bibitem{Adami:2021sko}
H.~Adami, M.~M. Sheikh-Jabbari, V.~Taghiloo, H.~Yavartanoo and C.~Zwikel,
  \emph{{Chiral Massive News: Null Boundary Symmetries in Topologically Massive
  Gravity}}, \href{http://dx.doi.org/10.1007/JHEP05(2021)261}{\emph{JHEP}
  {\bfseries 05} (2021) 261},
  [\href{https://arxiv.org/abs/2104.03992}{{\ttfamily 2104.03992}}].

\bibitem{Odak:2023pga}
G.~Odak, A.~Rignon-Bret and S.~Speziale, \emph{{General gravitational charges
  on null hypersurfaces}},
  \href{http://dx.doi.org/10.1007/JHEP12(2023)038}{\emph{JHEP} {\bfseries 12}
  (2023) 038}, [\href{https://arxiv.org/abs/2309.03854}{{\ttfamily
  2309.03854}}].

\bibitem{Ashtekar:1981bq}
A.~Ashtekar and M.~Streubel, \emph{{Symplectic Geometry of Radiative Modes and
  Conserved Quantities at Null Infinity}},
  \href{http://dx.doi.org/10.1098/rspa.1981.0109}{\emph{Proc. Roy. Soc. Lond.
  A} {\bfseries 376} (1981) 585--607}.

\end{thebibliography}
\end{document}